\documentclass{aastex631}

\usepackage{graphicx}

\newcommand{\mfd}{\texttt{lalpulsar\_Makefakedata\_v4}}
\newcommand{\cfs}{\texttt{lalpulsar\_ComputeFstatistic\_v2}}
\newcommand{\lalPredictFstat}{\texttt{lalpulsar\_PredictFstat}}
\newcommand{\fstat}{$\mathcal{F}$-statistic}
\newcommand{\twof}{2\mathcal{F}}

\newcommand{\Tobs}{T_{\mathrm{obs}}}
\newcommand{\dee}[1]{d#1\,}
\usepackage{amsmath}
\usepackage{amssymb}
\newcommand{\matr}[1]{\mathbf{#1}}

\newcommand{\Hz}{\ \mathrm{Hz}}

\begin{document}
\title{Analysing  one- and two-bit data to reduce memory requirements for
$\mathcal{F}$-statistic--based gravitational wave searches}

\author{P. Clearwater}
\affiliation{OzGrav, University of Melbourne, Parkville, Victoria 3010, Australia}
\affiliation{Data61, Commonwealth Scientific and Industrial Research Organisation,\\Corner Vimiera and Pembroke Roads, Marsfield, NSW 2122, Australia}
\author{A. Melatos}
\affiliation{OzGrav, University of Melbourne, Parkville, Victoria 3010, Australia}
\author{S. Nepal}
\affiliation{Data61, Commonwealth Scientific and Industrial Research Organisation,\\Corner Vimiera and Pembroke Roads, Marsfield, NSW 2122, Australia}
\author{M. Bailes}
\affiliation{OzGrav, Swinburne University of Technology, Hawthorn, Victoria 3122, Australia}

\begin{abstract}
	Searches for continuous-wave gravitational radiation in data collected by
	modern long-baseline interferometers, such as the Laser
	Interferometer Gravitational-wave Observatory (LIGO), the Virgo
	interferometer and the
	Kamioka Gravitational Wave Detector
	(KAGRA), can be
	memory
	intensive. A digitisation scheme is described that reduces the 64-bit
	interferometer output to a
	one- or two-bit data stream while minimising distortion and achieving
	considerable reduction in storage and input/output cost.
	For the representative
	example of the coherent, maximum-likelihood matched filter known as the
	\fstat{}, it is found
	using Monte-Carlo simulations
	that the injected signal only needs to be $\approx 24$ per cent stronger (for
	one-bit data) and $\approx 6.4$ per cent stronger (for two bit data with optimal
	thresholds) than a 64-bit signal in order to be detected with 90 per cent
	probability in Gaussian noise. The foregoing percentages do not change significantly when the
	signal frequency decreases secularly, or when the noise statistics are not
	Gaussian, as verified with LIGO Science Run 6 data.
\end{abstract}

\section{Introduction}
A key class of targets for long-baseline gravitational-wave interferometers,
such as the Advanced Laser Interferometer Gravitational-wave Observatory
(aLIGO) \citep{2015lscadvancedligo}, Virgo \citep{2015acernese} and the Kamioka
Gravitational Wave Detector (KAGRA) \citep{2021akutsu}, are persistent, quasi-monochromatic sources, called continuous-wave sources.
An important class of continuous-wave emitters are neutron stars, which radiate
at simple multiples of the stellar spin frequency $f_\star$.
Examples include
the mass quadrupole radiation from quasistatic
mountains (at $f_\star$ and $2f_\star$) \citep{2005melatospayne,2000ushomirsky},
current
quadrupole radiation from $r$-modes (at
approximately $4f_\star/3$) \citep{bondarescu2009,2015idrisy}, or non-axisymmetric flows in superfluids with pinned vortices
(at $f_\star$) \citep{2015melatosdouglass}.
For a source of known spin frequency, a common search technique
is to apply a coherent matched filter based on a maximum likelihood estimator called the \fstat{} \citep{1998jks}.
If the spin
frequency is unknown, semi-coherent search techniques are often preferred.
In this
paper
we consider specifically the coherent \fstat{}, but semi-coherent
techniques such as
the hidden Markov model \citep{2016suvorova,2017suvorova,2021melatos},
cross-correlation \citep{2008dhurandhar,2011chung,2018meadors},
PowerFlux \citep{2012dergachev,2013dergachev},
StackSlide \citep{2000brady,2009pletsch,2010pletsch,2015wette,2016wette,2018wette},
TwoSpect \citep{2011goetz,2016meadors} and the
Hough transform \citep{2004krishnan,2014astone_freqhough}
may also benefit from the ideas presented.

One challenge faced by gravitational wave data analysis is the large volume of
data involved.
Typically, a continuous-wave search processes 30-minute short Fourier
transforms (SFTs), sampled typically at kilohertz rates.
Data in SFTs are normally stored using 64 bit values for each sample.
Each SFT file
consumes $14.4$
kilobytes per $1\,\mathrm{Hz}$ of bandwidth.
For a typical year-long run,
this translates to approximately 250 gigabytes, per detector, to cover a $1\,\mathrm{kHz}$
band.
While this is not
a problem to store at present, a search that processes the whole dataset in a
random-access manner
[such as a cross-correlation--style
search \citep{2011chung}] can be slowed drastically due to
input/output (I/O) overhead when
processing more data than can be stored in random access memory (RAM).
Moreover, as search algorithms and data collection practices evolve, and new
generations of interferometers are constructed, there may come a time in the
future when memory limitations pose greater challenges than one envisions now,
as occurred historically with the move to baseband recording in pulsar radio
astronomy, e.g. \citet{2000stairs}.

In this paper, we show that \fstat{}-based searches work almost as well on one-
and two-bit digitisations of the original 64-bit data.
This
approach has been exploited in other low--signal-to-noise searches for periodic
signals in astronomy, such as radio pulsar timing and search experiments
\citep{1996manchester,1998jenet}.
The paper is structured as follows. In Section~2, we discuss and quantify the
distortion
introduced by digitisation. In Section~3, we review the \fstat{} and define the
multiple
digitisation schemes we investigate. In
Section~4, we report the results of Monte-Carlo performance tests on the
various
digitisation schemes, with and without frequency evolution.
In Section~5, we investigate how the results change, when the noise statistics
are not Gaussian.
In Section~6, we discuss briefly the path to implementing
the technique in existing software libraries for gravitational wave data
analysis.

\section{Optimal digitisation}
\label{sec:digitisation}
Storing an analogue signal as a digital file always involves a choice of how to
encode the data. The standard choice for science data collected from
gravitational wave interferometers is to store strain data in files called
frame files, as 
Institution of Electrical and Electronics Engineers standard
754 \citep{ieee754}
64-bit double-precision floating point numbers \citep{LIGO-T970130}.
When these frame files are converted into SFTs, the SFT files store two
single-precision (32-bit) floating point numbers for each frequency bin,
corresponding to the real and imaginary components, giving
an effective 64-bits of precision \citep{SFTformat}.
The error introduced by
digitisation to 64 bits is small compared to the
error arising from filtering in the interferometer's analogue-to-digital (ADC) converter, as
long as the ADC output exceeds about 13 bits, which suggests that 64 bits per
bin in the SFT files may be an unnecessary level of precision.
For a detailed
discussion of ADC digitisation noise, see \citet{LIGO-T970128-00-E}.

It is a counter-intuitive fact that signal processing algorithms operating
under low
signal-to-noise conditions suffer remarkably little degradation in sensitivity
when the data are truncated from 64 bits to (say) one or two bits.
Qualitatively,
this happens because one faces rapidly diminishing returns from greater
precision: there is no point measuring detector output precisely if the signal
is weak, because most of that
effort goes into measuring the noise.
The search for continuous gravitational waves
from neutron stars
is an example of such a low
signal-to-noise case.
Furthermore the signal-to-noise ratio stays low throughout the observation,
because the signal has
constant amplitude.
(One may compare
to radio pulsars, where focusing by the interstellar medium can temporarily
magnify
the signal-to-noise ratio, in which case digitisation
costs sensitivity.)
Digitisation can also improve the
robustness of a search by filtering out short-duration glitches.

The general theory of waveform distortion and sensitivity loss through
digitisation
was laid out in detail by \citet{1976max} and subsequently applied to radio
pulsar astronomy by \citet{1998jenet}. We summarise the main points in this
section to provide a framework for the application that follows. We refer the
reader to the aforementioned references for a fuller treatment of the subject.

Consider a continuous, real-valued time series $x(t)$. Suppose we construct a
digitised version of $x(t)$, denoted by $\hat{x}(t)$, by mapping the $N$
discrete intervals $[x_i, x_{i+1})$ ($i=1, ..., N$) to the $N$ discrete values
	$y_i$. Specifically, we have $\hat{x}(t) = y_i$ for $x_i \leq x(t) < x_{i+1}$
	at every sampled instant $t$.
	The first and last endpoints are necessarily $x_1 = -\infty$ and $x_{N+1} =
+\infty$ to ensure the digitiser covers the full input range.
The task is to minimise
the distortion introduced by digitisation,
which is equivalent to minimising the expected value
\begin{equation}
	\rho^2 = \left\langle[x(t) - \hat{x}(t)]^2\right\rangle,
	\label{eqn:evdistortion}
\end{equation}
where $x(t)$ is a random variable equal to the signal plus the noise\footnote{The
	mean-square error in equation~(\ref{eqn:evdistortion}) is not the only valid
measure of distortion of course, but it is favoured in the literature
\citep{1976max,1998jenet}.}. Equation~(\ref{eqn:evdistortion}) is minimised with
respect to $x_2, ..., x_N$, yielding the $N-1$ simultaneous equations
\begin{equation}
	x_i = (y_{i-1} + y_i)/2,  \qquad \mathrm{for\ } i = 2, ..., N,
		\label{eqn:optthresh1}
\end{equation}
and also minimised with respect to $y_1, ..., y_N$, yielding the $N$ simultaneous equations
\begin{equation}
		0 = 
		\int_{x_i}^{x_{i+1}} \dee{x} (x-y_i)p(x), \qquad
	\mathrm{for\ } i = 1, ..., N,
		\label{eqn:optthresh2}
\end{equation}
where $p(x)$ is the probability density function of the sampled time series
$x(t)$.

In what follows, we solve Equations~(\ref{eqn:optthresh1})
and~(\ref{eqn:optthresh2}) for Gaussian noise with standard deviation $\sigma$
in the low signal-to-noise
regime, i.e., $p(x) \approx (2\pi\sigma)^{-1/2}\mathrm{exp}(-x^2/2\sigma^2)$.
The results are given in Section~\ref{sec:searches}.
Strictly
speaking, for gravitational wave applications,
it is better to minimise the distortion of the \fstat{} itself,
rather
than the input time series, but
the results reported in Section~\ref{sec:searches} are excellent in practice
without this hard-to-calculate refinement, so we defer it to future work.

\section{Few-bit $\mathcal{F}$-statistic searches}\label{sec:searches}
\subsection{\fstat}
The \fstat{} \citep{1998jks} is a maximum-likelihood estimator for the
log likelihood
of detecting a gravitational wave signal at a specific frequency (and its time
derivatives) and sky position in noisy interferometer data. The \fstat{}
accounts for modulation in the signal caused by the
Earth's rotation and its orbital motion about the Sun.
An explicit recipe for calculating $\mathcal{F}$ from the interferometer output
(in the form of Fourier transforms), antenna pattern functions, noise spectral density, and
phase model is given in Section 3 of \citet{1998jks}
and is not reproduced here.

Given stationary, Gaussian noise with no signal, $\twof$ is distributed
according to a chi-squared distribution with four degrees of freedom,
whose probability density function (PDF) we denote by $\chi_4^2(\lambda = 0)$,
where
$\lambda$ is the
non-centrality parameter (zero when there is no signal).
Given a desired false alarm
probability $\alpha$, the formula
\begin{align}
	\alpha = \int_{2\mathcal{F}_c}^{\infty} \dee{(2\mathcal{F})} \chi^2_4(\lambda = 0;
	2\mathcal{F})
	\label{eqn:digit:2Fc}
\end{align}
defines a detection threshold $2\mathcal{F}_c$.

In the presence of a signal, the \fstat{} is distributed according to a
non-central $\chi^2$ distribution (again with four degrees of freedom), where the non-centrality parameter $\lambda$
depends on the dimensionless wave strain $h_0$, the one-sided detector noise
spectral density $S_h(f)$ at the observing frequency $f$, and the
observation time $\Tobs$ according to
\begin{align}\label{eqn:lambdah0}
	\lambda = \frac{Ah_0^2\Tobs}{S_h(f)}.
\end{align}
The constant of proportionality $A$ depends on the various parameters
describing the
source and detector orientation.
If some of those parameters are unknown,
they can be averaged over
\citep{1998jks}.
Given a desired false dismissal
probability $\beta$, the condition
\begin{align}
	\label{eqn:beta}
	\beta = \int^{2\mathcal{F}_c}_{0} \dee{(2\mathcal{F})} \chi^2_4(\lambda; 2\mathcal{F})
\end{align}
implicitly defines a squared signal-to-noise ratio $\lambda$ corresponding to
the
threshold $2\mathcal{F}_c$.

\begin{figure}%
	\includegraphics[width=0.49\columnwidth]{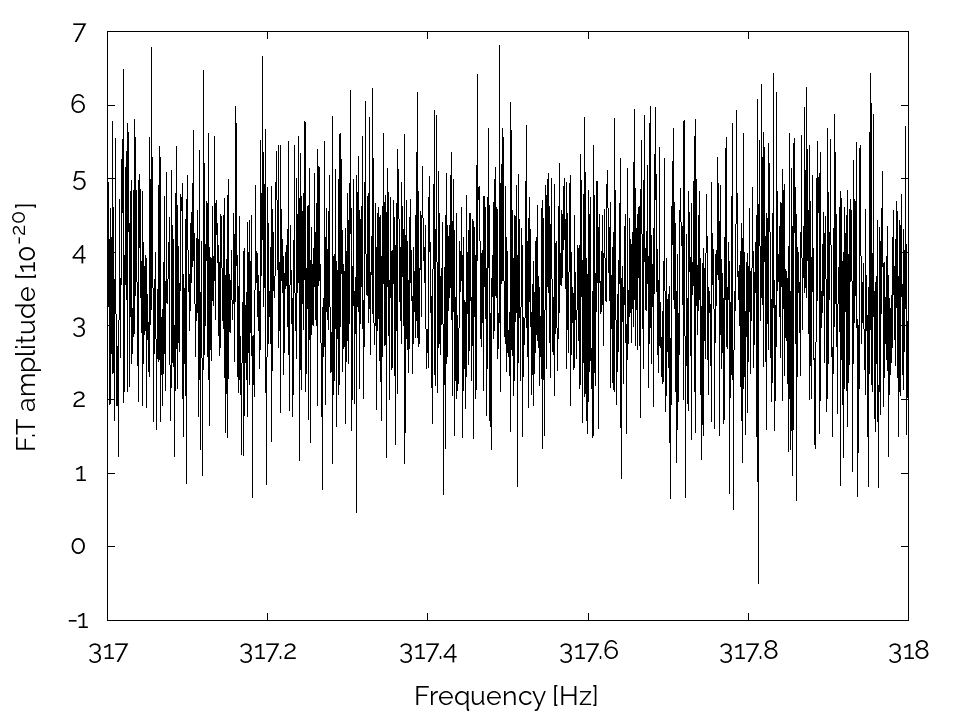}
	\includegraphics[width=0.49\columnwidth]{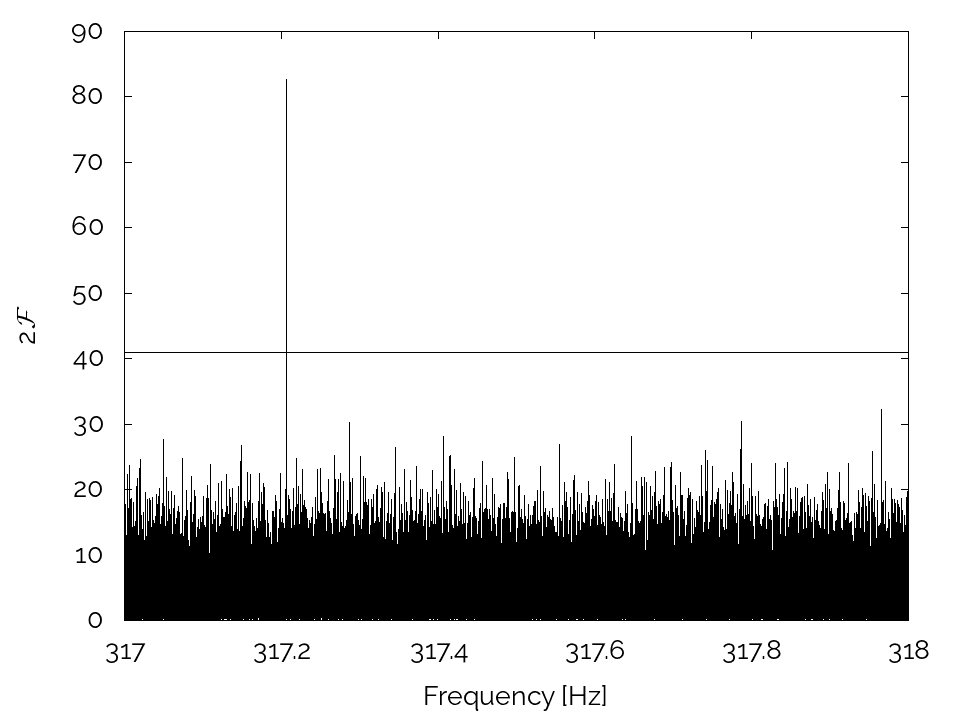}
	\includegraphics[width=0.49\columnwidth]{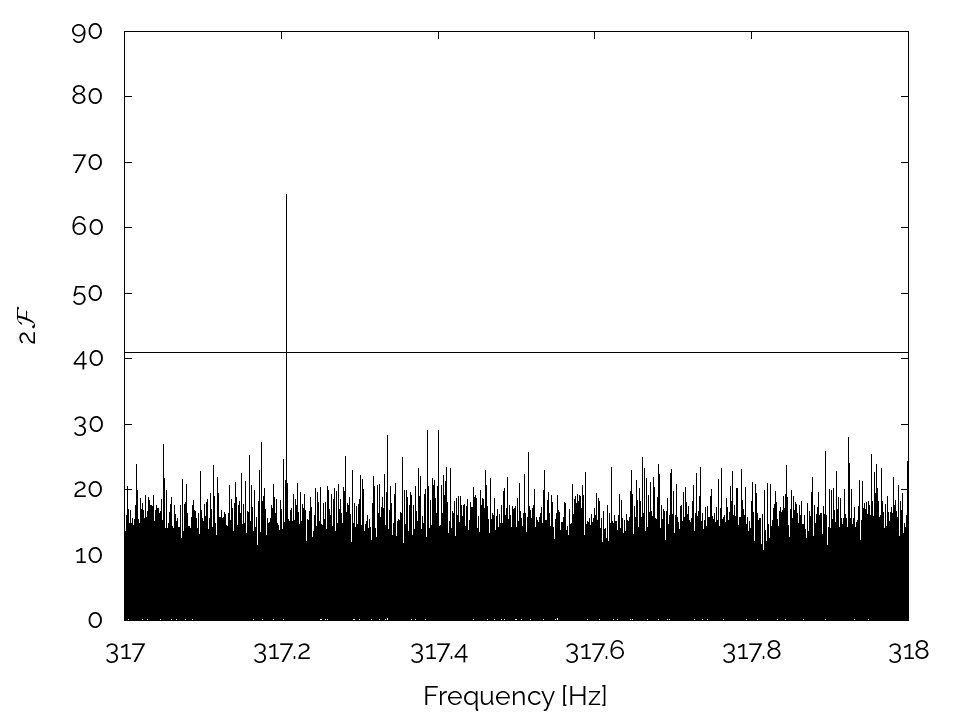}
	\includegraphics[width=0.49\columnwidth]{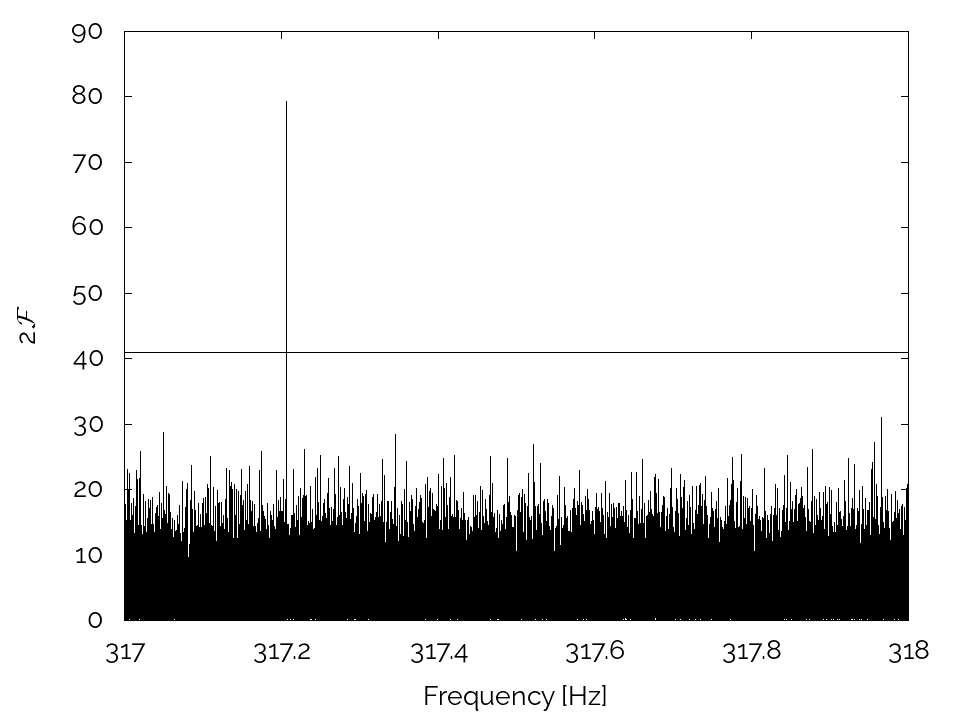}
	\caption
	{
		Representative plot of the search output from one noise realisation,
		including
		an injected signal with the parameters in Table~\ref{tbl:simparams} and
		a signal strength of
		$h_0 = 6\times10^{-25}$.
		The horizontal axis is the observing frequency.
		The top left plot displays the summed squared modulus of the 96
		short Fourier transforms in the data set; no signal is visible. The top right
		plot displays the
		\fstat{} with full 64-bit digitisation; the signal is visible at
		$317.26\,\mathrm{Hz}$. The bottom left and bottom right
		plots display the \fstat{} with one- and two-bit digitisation
		respectively.
		The horizontal bar is the critical
		threshold $2\mathcal{F}_c=40.96$.
	}
	\label{fig:fiducialfstat}
\end{figure}

If a search is done over $N$ statistically independent
measurements, the total false alarm probability is
\begin{equation}
	\label{eqn:alphaN}
	\alpha_N = 1 - \left[\int_{0}^{2\mathcal{F}_c} \dee{(2\mathcal{F})} \chi^2_4(\lambda = 0;
2\mathcal{F})\right]^N.
\end{equation}
In the tests performed in the next section,
we search over a band spanning $1.06 \Hz$ for an observing time of
$T_{\mathrm{obs}} = 48$ hours.
The LIGO--Virgo--KAGRA Algorithms Library (LALSuite) \citep{lalsuite} \fstat{} implementation,
\cfs{} (formerly known as \texttt{lalapps\_ComputeFStatistic\_v2}),
takes the frequency spacing as a parameter that zero-pads the time series,
which enables the user to select a frequency bin size.
The frequency spacing is sometimes related colloquially to
an over-resolution parameter, $r$, in the literature \citep{2014sammut},
although it does not correspond necessarily to over-resolution in the strict
sense.
A typical choice (and incidentally the default value), which
we follow here, is to space the bins by $1/(2T_{\mathrm{obs}})$ (i.e. $r=2$), used
for example in
many
published continuous wave searches
\citep{2015aasiScoX1Initial,2017abbottScoX1HMM,2019abbottV2O2,2020middleton,2020millhouse,2021beniwal,2022lsc_O3ScoX1HMM,2022lsc_julianlmxb}.
Hence, the number of \fstat{} frequency bins is 
$N_{\mathrm{bins}} = 1.06\,\mathrm{Hz} \times 172\,800\,\mathrm{s} \times 2 =
366\,336$, and the frequency bin size is
$1/(2 \Tobs) = 2.89\times10^{-6}\,\mathrm{Hz}$.

In passing we note that the assumption of statistical independence in the
previous paragraph is often not true: when searching over a region of parameter
space, search points that are close in parameter space have similar
parameters, and the \fstat{}
is therefore correlated.
Assuming instead that the \fstat{} is independent is
conservative (causing the false alarm probability to be overstated) and
so is favoured in the literature, although more scrutiny is
being applied to the assumption recently \citep{2022tenorio}. For this
paper, we adopt the assumption of statistical independence. In the tests
described below, we check the actual false alarm rate to ensure that it is at
the level expected.

For the purposes of the tests described in the rest of this paper, we choose
$\alpha_N = 0.01$ and $\beta
= 0.1$.
Solving Equations~(\ref{eqn:beta}) and (\ref{eqn:alphaN}) with these probabilities
yields $2\mathcal{F}_c =
40.96$.
These choices of false alarm and false dismissal probability are typical for
continuous gravitational
wave searches
\citep{2005lsc-hough,2005lsc-pulsarlimits}, although we note that $N$ varies
widely amongst searches.

\subsection{Digitisation scheme}
\label{ssec:digschemes}
We test
the performance of the \fstat{} for several one- and two-bit digitisation
schemes. Besides the optimal scheme described in Section 2, we examine some
suboptimal but intuitively natural schemes to test whether or not the results
are sensitive to how the digitisation intervals and output values are selected.

For one-bit digitisation, we simply apply the sign function to the
input signal:
\begin{align}
	\label{eqn:one-bit-sgn}
	\hat{x}(t) = \begin{cases}
		+1, \qquad \mathrm{if}\ x(t) \geq 0, \\
		-1, \qquad \mathrm{if}\ x(t) < 0.
	\end{cases}
\end{align}
The means of $x(t)$ and $\hat{x}(t)$ are zero, so Equation~(\ref{eqn:one-bit-sgn})
does not introduce a DC bias.
The \fstat{} value is independent of the total input power, so we do not need to
normalise $\hat{x}(t)$.

For two-bit digitisation, we apply the four-level function
\begin{equation}
	\hat{x}(t) = \begin{cases}
		-3, \qquad & x_1 \leq x(t) < x_2, \\
		-1, \qquad & x_2 \leq x(t) < x_3, \\
		+1, \qquad & x_3 \leq x(t) < x_4, \\
		+3, \qquad & x_4 \leq x(t) < x_5. \\
	\end{cases}\label{eqn:spacing}
\end{equation}
Again, we do not normalise $\hat{x}(t)$. We do choose arbitrary output values
that
are equally spaced, a point to which we return below.

We test three approaches to choosing the thresholds $x_1, ..., x_5$.
For a zero-mean signal in Gaussian noise, there is no loss of generality in
choosing
$x_1 = -\infty$, $x_3 = 0$ and $x_5 = +\infty$.
As the PDF of $x(t)$ is
symmetric, we also have $x_2 = -x_4$, leaving $x_4$ to be specified.

In the first approach, we set $x_{4} = \mathrm{max}|x(t)|/2$, where the maximum
is computed empirically from the whole measured time series.
The drawback is that the
maximum is usually a many-sigma outlier, so few
samples
are actually digitised to the $\pm3$ levels.

In the second approach, we seek to map
equal numbers of
samples to each output level. We set $x_3$ to equal the
median sample to create two
equal-size partitions and set $x_2$ and $x_4$ to the medians of
the two partitions. This recipe does not guarantee $x_3 = 0$ and $x_2 = -x_4$
exactly, but
in practice these relations hold to a good approximation for large data sets.
Note that substantial
pre-processing is needed to determine the quartiles.

A third approach follows the methodology developed in
\citet{1976max}, and summarised in Section~\ref{sec:digitisation},
for choosing
optimal thresholds.
For a Gaussian
distribution with standard deviation $\sigma$, equations~(\ref{eqn:optthresh1})
and~(\ref{eqn:optthresh2}) with $N = 4$ can be solved to yield $x_1 = -\infty$,
$x_2 = -0.9816\sigma$, $x_3 = 0$, $x_4 = 0.9816\sigma$ and $x_5 = \infty$.
The corresponding output levels are $y_1 = -1.510\sigma$, $y_2 = -0.4528\sigma$,
$y_3 = 0.4528\sigma$, $y_4 = 1.510\sigma$.
The optimal output levels are not equally spaced, as in
equation~(\ref{eqn:spacing}), but they are within 15 per cent of being so. As
the \fstat{} is independent of input power, $\sigma$ may be chosen arbitrarily
for our purposes in $y_1, ..., y_4$ (but not in $x_1, ..., x_5$ of course).

\section{Performance tests with Gaussian noise}
\label{sec:tests}

\subsection{Monte-Carlo simulations}
\label{ssec:mc}
We compare the detection performance of the \fstat{} 
for one-, two-, and 
64-bit digitisation by
simulating multiple realisations of
the gravitational wave signal from an isolated, triaxial rotator with
spin frequency $f_0$
superposed on
stationary Gaussian noise.
The
fixed source and noise parameters
in each realisation are given in Table~\ref{tbl:simparams}.

We inject signals
with wave strains in the range $0.5\times10^{-25} \leq h_0 \leq
10.0\times10^{-25}$, in step
sizes of $0.5\times10^{-25}$.
The one-sided noise amplitude spectral density,
$\sqrt{S_h(2f_0)} = 2\times10^{-23}\,\mathrm{Hz}^{-1/2}$, is
chosen to be characteristic of the initial LIGO design.
Hence the chosen $h_0$ values
correspond to values of $h_0 [\Tobs/S_h(2f_0)]^{1/2}$ 
between $1.0$ and $20.8$. For each digitisation scheme, $200$ realisations are
created for each $h_0$ value to measure the detection probability.

The synthetic noise and signals are generated using the LALSuite
\citep{lalsuite} tool \mfd{} (formerly known as
\texttt{lalapps\_Makefakedata\_v4}).
The tool produces
either a time series or a sequence of SFTs.
In this paper we run two kinds of tests. Sometimes we digitise the time
series, whereupon we use \mfd{} to produce the time series, apply the
desired digitisation scheme, create the SFTs by using FFTW \citep{FFTW05}
and output them in the SFTv2 format
\citep{SFTformat}.
Sometimes we digitise the SFTs, whereupon we use \mfd{} to
produce SFTs, and then separately digitise the real and imaginary values to
produce new, digitised SFTs.

The resulting files are in the same format as standard SFTs, which means
that they can be consumed directly by LALSuite tools. However, in order to realise
the savings of digitisation, a new SFT in-memory format would have to be
developed, with LALSuite tools modified to read and write to that
format.

The search is performed using the LALSuite tool \cfs{}
\citep{prixTechnicalNote}, and
is run with the search parameters (other than frequency) exactly matching the injection
parameters, to ensure no loss in signal strength due to a mismatch between
search and injection parameters. As explained above, we search a
$1.06\,\mathrm{Hz}$ band containing the injected signal, thus searching
over $N_{\mathrm{bins}}$ choices of frequency. By doing this search (rather
than just calculating the \fstat{} for the injection bin), we mimic a real
search which typically searches frequency
over sub-bands $\sim 1\,\mathrm{Hz}$ wide
\citep{2017abbottScoX1HMM,2019abbottV2O2},
and use the band to set a detection threshold through
Equation~(\ref{eqn:digit:2Fc}). Of course, a real search may search over
additional search parameters,
or a narrower sub-band, if the frequency is measured electromagnetically with
high accuracy:
the choice to search over frequency
in these tests is for illustration.
We claim a detection if the bin containing the
frequency of the injected signal returns $2\mathcal{F} >
40.96$, corresponding to a one per cent false alarm rate.

In claiming a detection, we consider the value of $2\mathcal{F}$ only in the
bin containing the injection frequency and so disregard other bins, even those
over the threshold $2\mathcal{F}_c$. An alternative approach would be to claim a
detection when any bin in the search band exceeds the threshold, which of
course is the only option available in an astrophysical search, when the signal
frequency is often unknown. The principal aim of this paper is to compare different
digitisation methods to full-precision processing, rather than to quantify the
absolute sensitivity of a search, so for simplicity we take the approach of
looking only in the injection bin. For the Monte Carlo simulations with Gaussian
noise, we nevertheless monitor the whole search band for values that exceed
$2\mathcal{F}_c$
(other than those forming part of the signal peak),
to assure that the false alarm rate does not
exceed $\alpha = 0.01$.

Weak sidelobes are observed in the \fstat{} output at $f_0\pm9\times10^{-6} \,\mathrm{Hz}$
and $f_0\pm3.5\times10^{-5}\,\mathrm{Hz}$, whose origin is uncertain but likely
due to the diurnal motion of the Earth, which is not completely accounted for by
the \fstat{} due to approximations in its implementation
\citep{2007prix,prixTechnicalNote}.
They are not at the diurnal frequency $1/(86\ 400\,\mathrm{s}) =
1.16\times10^{-5}$, or a multiple thereof, perhaps due to an interaction with
the $1800\,\mathrm{s}$ SFT length.
The sidelobes are not germane to what follows.

\begin{table}
	\caption
	{Fixed source parameters for Monte-Carlo simulations.}
	\begin{center}
	\begin{tabular}{llll} \hline\hline
		Variable & Symbol & Value & Units \\ \hline 
		Reference GPS start time & & $875\,936\,746$ & s \\
		Observation time & $\Tobs$ & $172\,800$ & s \\
		Right ascension & $\alpha$ & $0.092408204211$ & rad \\
		Declination & $\delta$ & $-0.956660382019$ & rad \\
		Inclination angle & $\cos\iota$ & $-0.987561752324$ & \\
		Polarisation angle & $\psi$ & $0.423734950335$ & rad \\
		Initial phase & $\Phi_0$ & $3.691030880249$ & rad \\
		Spin frequency & $f_0$ & $317.207201716543$ & Hz \\
		Noise amplitude spectral density & $\sqrt{S_h(2f_0)}$ & $2\times10^{-23}$ &
		$\mathrm{Hz}^{-1/2}$ \\
		\hline
	\end{tabular}
	\end{center}
	\label{tbl:simparams}
\end{table}

\subsection{Detection probability}
\label{ssec:detrate}
Figure~\ref{fig:fiducialfstat} compares the output of the \fstat{} for one-,
two- and 64-bit digitisation for one representative realisation with
$h_0 = 6\times10^{-25}$ and $f_0$ as given in Table~\ref{tbl:simparams}. The first panel
shows the summed output (squared modulus) of the 96 SFTs for the full
simulated observation.
No signal is discernible at $f_0$. The second panel shows the
\fstat{} plotted versus frequency with full 64-bit digitisation. The signal is
now clearly apparent as a spike at $f_0$, with $2\mathcal{F} = 82.6$. The third
and fourth panels show the \fstat{} output with one- and two-bit digitisation
(optimal scheme, see Section~\ref{ssec:digschemes}).
The results are remarkably similar to the second panel. The
signal is clearly detected in both the one- and two-bit cases, with
$2\mathcal{F} = 65.0$ and $79.3$ respectively. In other words, digitisation
causes a modest and tolerable loss of \fstat{} power.

\begin{figure}
	\includegraphics[width=\columnwidth]{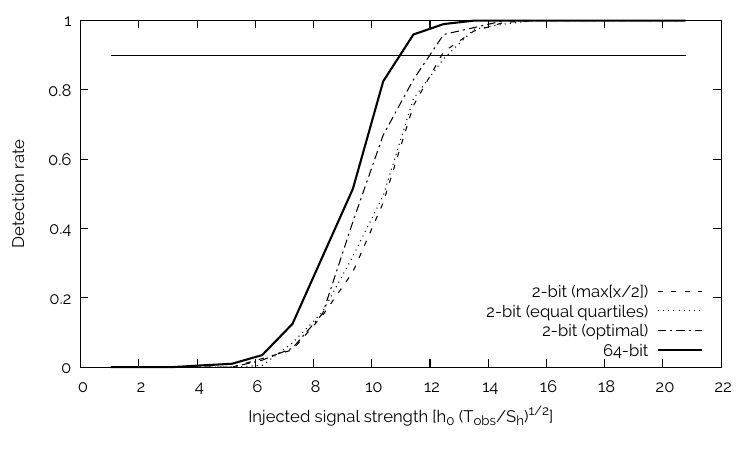}
	\caption
	{
		\fstat{} detection fraction versus injected signal-to-noise ratio
		$h_0[\Tobs/S_h(f_0)]^{1/2}$ for 64-bit time series (control) and three
		choices of two-bit digitisation: using $\mathrm{max}|x(t)|/2$ as a
		threshold, using four equal-size quartiles, and using the optimal
		thresholds (see legend). The detection probability is computed from 200 realisations,
		with $2\mathcal{F}(f_0) > 40.96$ counting as a detection from
		Equation~(\ref{eqn:alphaN}) with $\alpha_N = 0.01$.
		The horizontal line
	corresponds to 90 per cent detection probability, that is, a false
	dismissal probability of $\beta = 0.1$.
}\label{fig:twobit}
\end{figure}

Figure~\ref{fig:twobit} compares two- and 64-bit digitisation for the three
approaches to two-bit digitisation described in Section~\ref{ssec:digschemes}.
It plots the fraction of realisations that yield a successful detection
($2\mathcal{F} >
2\mathcal{F}_c$) versus $A^{-1}$ times the signal-to-noise ratio $\lambda$, i.e., $h_0
[\Tobs/S_h(2f_0)]^{1/2}$.
Digitisation does not affect the overall shape of the curve. As
expected, the 64-bit analysis leads to the highest detection probability. Two-bit
digitisation reduces the detection probability modestly at all signal-to-noise
ratios. Of the digitised
curves,
choosing $x_4 = \mathrm{max}|x(t)|/2$ leads to the
poorest detection performance, but even then the recovered
\fstat{} value
drops by just 11 per cent relative to a full 64-bit analysis.
Equally weighted quartiles yield intermediate performance, and the optimal
thresholds deliver, unsurprisingly, the best two-bit performance.
The reduction in \fstat{} value
for optimal thresholds is
6.4 per cent
relative to a 64-bit analysis.

\begin{figure}
	\includegraphics[width=\columnwidth]{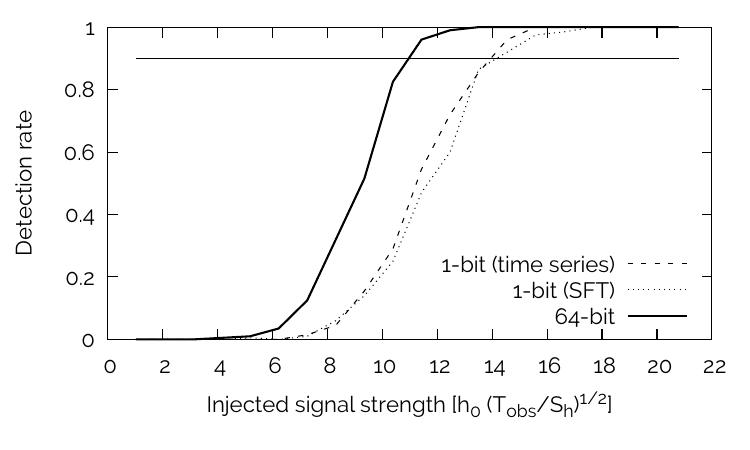}
	\caption
	{
		\fstat{} detection fraction versus injected signal-to-noise ratio
		$h_0[\Tobs/S_h(f_0)]^{1/2}$ for 64-bit time series (control), one-bit
		time series, and one-bit SFTs (see legend).
		The detection probability is computed from 200 realisations,
		with $2\mathcal{F}(f_0) > 40.96$ counting as a detection from
		equation~(\ref{eqn:alphaN}) with $\alpha_N = 0.01$. The horizontal line
	corresponds to 90 per cent detection probability.
}
\label{fig:tsvssft}
\end{figure}

Figure~\ref{fig:tsvssft} compares full 64-bit digitisation against
one-bit digitisation of the time series and the SFT.
When digitising the SFT, the resulting \fstat{} exhibits a different PDF from
the expected chi-squared distribution, as discussed further in
Section~\ref{subsec:digitisation:PDF}. We empirically determine that the critical
threshold to achieve a false alarm probability of $\alpha_N = 0.01$ with one-bit digitised SFTs
equals $2\mathcal{F}_c = 26.09$.
As
with two-bit digitisation, the shapes of all three curves are similar;
essentially they are horizontally shifted relative to one another.
Digitisation of
the time series ($24$ per cent sensitivity loss) is marginally more effective than
digitisation of the
SFTs ($29$ per cent sensitivity loss).

The intersection between the horizontal line indicating a 90 per cent detection
probability and each curve in Figures~\ref{fig:twobit} and~\ref{fig:tsvssft}
indicates how much stronger a signal must be in order to be detected in few-bit
digitised data.
These values are summarised, for the representative case of $\alpha_N = 0.01$,
$\beta = 0.9$ in
Table~\ref{tbl:magic-numbers} for the six digitisation schemes considered.
While the sensitivity losses listed in Table~\ref{tbl:magic-numbers}
cannot be ignored, it is possible to recover full sensitivity by using a one- or
two-bit digitised search as a first pass with a lower threshold (i.e. higher
$\alpha_N$) to identify
likely signals. Those candidates can then be reprocessed using the full
precision data.

\begin{table}
	\caption
	{Strength of an injected signal necessary
	to achieve 90 per cent detection efficiency at $\alpha_N = 0.01$, as a
	function of digitisation
	scheme. The efficiency is determined empirically from $10^3$ Monte Carlo realisations.}
\label{tbl:magic-numbers}
\begin{center}
\begin{tabular}{ll} \hline\hline
	Digitisation scheme & $h_0 \sqrt{\Tobs/S_h}$ \\ \hline
	64-bit & 11.08 \\
	2-bit, time series, optimal & 11.79 \\
	2-bit, time series, $\mathrm{max}|x(t)|/2$ & 12.57 \\
	2-bit, time series, equal quartiles & 12.58 \\
	1-bit, time series & 13.78 \\
	1-bit, SFTs & 14.30 \\
	\hline
\end{tabular}
\end{center}
\end{table}

\subsection{PDF of \fstat{} applied to digitised data}
\label{subsec:digitisation:PDF}
\begin{figure*}
	\begin{center}
	\includegraphics[width=0.49\columnwidth]{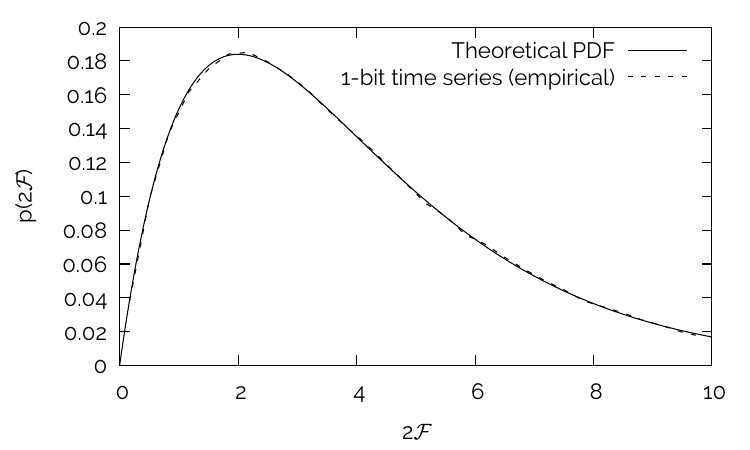}
	\hfill
	\includegraphics[width=0.49\columnwidth]{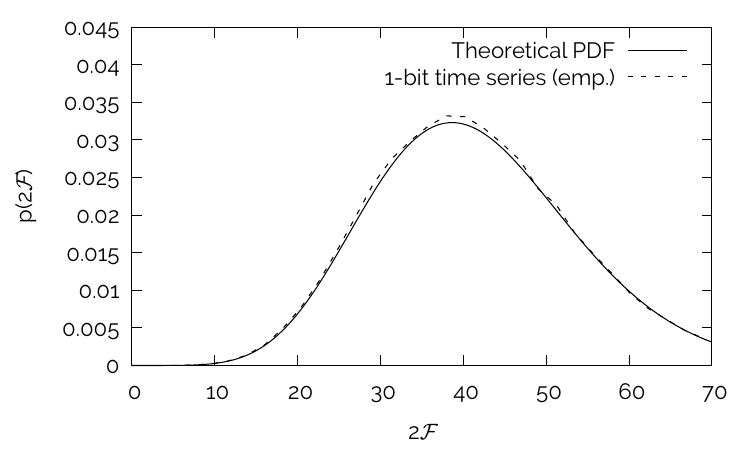}
	\hfill
	\includegraphics[width=0.49\columnwidth]{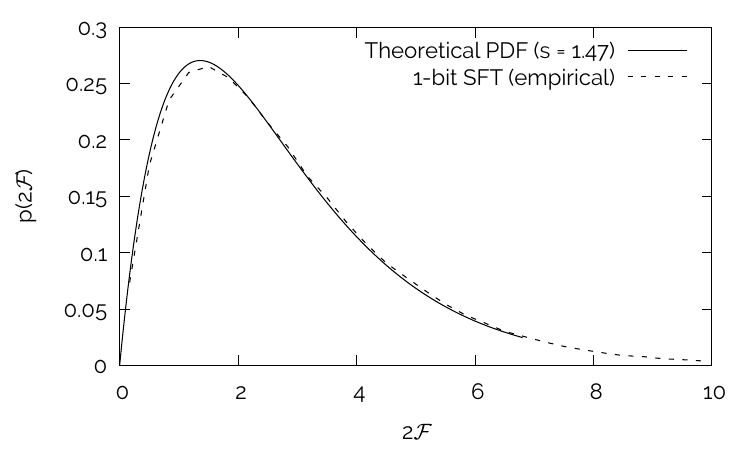}
	\hfill
	\includegraphics[width=0.49\columnwidth]{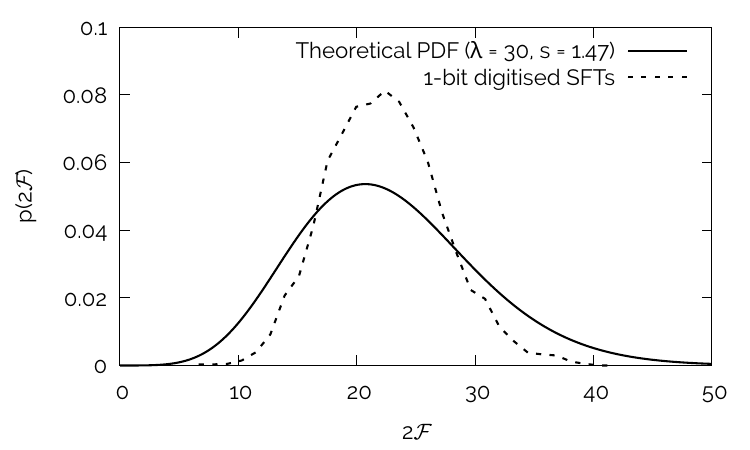}
	\hfill
	\end{center}
	\caption
	{
		Probability density function (PDF) for the \fstat{} operating on
		one-bit digitised data. The left column shows noise-only realisations,
		while the right column shows noise plus an injected signal. The top row
		shows digitisation of the time series, and the bottom row shows
		digitisation of the SFTs.
		In all panels, the dotted curve is the empirical PDF derived from Monte
		Carlo simulations with $10^5$ realisations.
		For the top row, the solid curve shows the theoretically expected PDFs
		for 64-bit digitisation, namely a chi-squared distribution (top left)
		and a non-central chi-squared distribution (non-centrality parameter
		$\lambda = 37$). The empirical distributions for the time-series
		digitised data are good fits to the theory, as discussed in
		Section~\ref{subsec:digitisation:PDF}. By contrast, the empirical PDFs
		in the bottom row (where the
		SFTs are digitised) are distorted significantly compared to the PDF for
		64-bit digitisation. In the noise-only case (bottom left panel), an
		\textit{ad hoc} scaling factor $s = 1.47$ produces a good match (see
		Section~\ref{subsec:digitisation:PDF}). However, no simple scaling can
		reproduce a chi-squared distribution when digitising data containing an
		injected signal (bottom right panel; non-centrality parameter $\lambda
		= 30$ chosen as best fit with scale factor $s = 1.47$).
	}\label{fig:probdist}
\end{figure*}

For completeness, we present the empirical PDF of the
\fstat{} after one-bit digitisation in Figure~\ref{fig:probdist}.
We find that the PDF is unchanged to an excellent approximation, relative to
double-precision processing, when the time
series is digitised to one bit; the discrepancy is not discernible by eye.

When
the SFTs are digitised,
as opposed to the time series,
the \fstat{} PDF does change (see histograms in bottom row) and exhibits a
thinner tail than its 64-bit counterpart.
The top left panel shows the
\fstat{} PDF in the noise-only case, with time series digitisation, which is
predicted theoretically to be $\chi^2_4(\lambda = 0; 2\mathcal{F})$ (solid
curve).
The
top right panel shows the \fstat{} PDF when a signal is present, again with
digitisation in the time series, which is
predicted to be 
$\chi^2_4(\lambda \neq 0; 2\mathcal{F})$ (solid curve).
The empirical PDFs (dashed curves) are
generated from $10^5$ realisations of the \fstat{} by selecting $2\mathcal{F}$ from a
frequency bin where the signal is absent (top left panel) and present (top
right and bottom panels).

The cause of the distortion in the PDF of the \fstat{}, when digitising the
SFT, is uncertain.
Empirically, when only noise is present, multiplying $2\mathcal{F}$ by a scale factor $s
= 1.47$ restores the PDF to the expected shape, that is, $2s\mathcal{F}$
(rather than $2\mathcal{F}$) is $\chi^2_4$ distributed.
The value of $s$ is not affected by the choice of $\Tobs$, the length of time covered
by each SFT, or the right ascension or declination searched.
When processing the data at progressively higher precision, the PDF converges on the
chi-squared distribution, becoming indistinguishable by eye at about five-bit
precision,
when using the output levels recommended by
\citet{1976max}.

In the bottom-right panel of Figure~\ref{fig:probdist} we plot the PDF of the \fstat{}
in the noise plus signal case,
operating
on one-bit digitised SFTs,
with the object of testing whether $2\mathcal{F}$ follows a non-central chi-squared
distribution in the presence of a signal.
The PDF is determined through Monte Carlo simulations
containing a strong injection, $h_0 = 2\times10^{-24}$, with other parameters
given in Table~\ref{tbl:simparams}.
Adopting the scaling factor $s = 1.47$ from the noise case, we fit a non-central
chi-squared distribution with four degrees of freedom.
This fit finds the best-fit non-centrality parameter $\lambda = 29.5$,
but the resulting curve is a poor fit to the empirical PDF, which is
narrower and more sharply-peaked.
This implies that the PDF distortion occurs for both noise and signal (i.e. it
is not restricted to one or the other), and that the scale factor $s$ does not
work for a signal even though it works for noise. These tests may help inform
future work to identify the cause of the distortion.

\subsection{Frequency evolution}

Isolated neutron stars spin down due to electromagnetic braking
\citep{1997melatos}.
The performance of a coherent
matched filter like the \fstat{} degrades, if the
source frequency evolves significantly over the observation period, unless the
frequency evolution is known and accounted for when computing the \fstat{}.
As the frequency evolution of the source is often not known precisely (for
example, because of uncertainties in electromagnetic measurements of spin
frequency, or because the frequency evolution incorporates stochastic terms),
it is important to compare the loss of sensitivity
caused by a changing source frequency between the 64-bit and fewer-bit cases.

\begin{figure}
	\includegraphics[width=0.99\columnwidth]{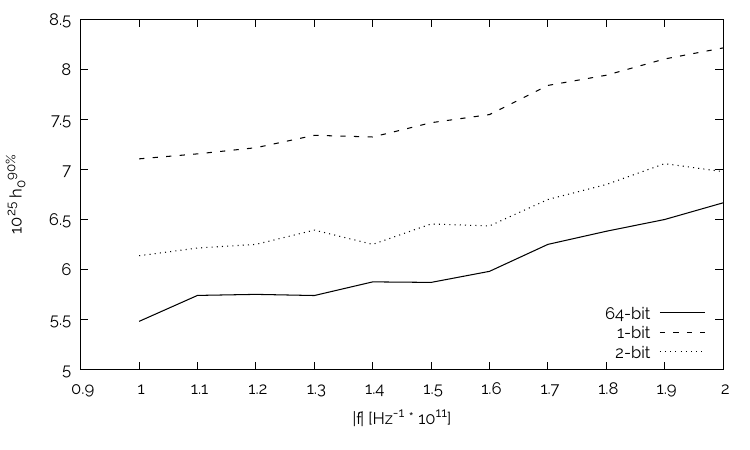}
	\caption
	{
Sensitivity loss caused by digitisation versus
	spin-down rate $\dot{f}$ (horizontal axis; units $10^{11}
	\mathrm{Hz}^{-1}$). The
	vertical axis gives the wave strain $h_0^{90\%}$ of the injected signal
	necessary for a 90 per cent probability of detection.
	As expected, as $|\dot{f}|$ increases, so does $h_0^{90\%}$, but the slope
	of the $h_0^{90\%}$ versus $|\dot{f}|$ curve is roughly the same for one-bit
	(dashed line), two-bit (dotted line)
	and 64-bit (solid line) digitisation.
		}
	\label{fig:fdot}
\end{figure}

For the purposes of this paper, we model the spin down by the leading term in
the Taylor expansion proportional to $\dot{f} \neq 0$ in the \mfd{} signal
phase while taking $\dot{f} = 0$ in the \fstat{}\footnote{In practice,
\cfs{} can search over $\dot{f}$, $\ddot{f}$
and $\dddot{f}$ but the problem then transfers to higher derivatives, which
degrade the sensitivity in an analogous fashion.}. \citep{1998jks}
We compare the percentage sensitivity loss with one-, two- and 64-bit digitisation
and two-bit digitisation (using optimal thresholds)
for $0 \leq \dot{f} \leq 2\times10^{-11}\ \mathrm{Hz}\,\mathrm{s}^{-1}$. For
$\dot{f} > 2\times10^{-11}\ \mathrm{Hz}\,\mathrm{s}^{-1}$ (with the
parameters selected in Table~\ref{tbl:simparams}), the signal cannot be
detected reliably even with 64 bits. The results are displayed in Figure~\ref{fig:fdot}.
The few-bit curves are parallel to the 64-bit curve for all $\dot{f}$, implying
that few-bit digitisation does not produce any \textit{extra} loss of sensitivity
due to $\dot{f} \neq 0$, over and above the loss of sensitivity measured in
Section~\ref{ssec:detrate}.

\section{Realistic noise}
\label{sec:nongauss}
The tests in Section~\ref{sec:tests} assume white, Gaussian noise. In this
section, we examine those two assumptions and use data collected by LIGO as background
noise to test how digitisation performs, when these assumptions are weakened.

Although the noise power spectrum of the LIGO detectors varies substantially
over its operating
band \citep{2015lscadvancedligo},
continuous-wave searches based on the \fstat{} are typically broken up into
multiple subsearches addressing narrow subbands (typically 1 Hz wide) covering
only a small part of the total operating band.
Therefore, we can safely take
the noise spectrum in each subband to be approximately flat (i.e. white noise).
The same is true when digitising the time series, as long as
the data are bandpass filtered (otherwise the digitisation tends to track
sinusoidal variation of large noise lines).

While Gaussian noise is often assumed for the purpose of benchmarking
different detection strategies \citep{1998jks}, in practice LIGO detector noise
is routinely non-Gaussian,
and varies depending on the detector, observation start
time, and frequency band \citep{2015aasi}.
It is beyond the scope of this paper to exhaustively analyse LIGO detector
noise; for an example of recent discussion of data quality in Advanced LIGO see
\citet{2018covas}.

Nonetheless, it is instructive to perform some representative
investigations of
whether digitisation remains effective with non-white and
non-Gaussian noise. We repeat the tests from Section~\ref{sec:tests} in two
ways:
we attempt to detect hardware injections made during LIGO's sixth science run
(S6),
and we attempt to detect synthetic signals injected on top of S6 data
in software.
In the latter case, we investigate the effect of varying
the sky position, search
frequency band, and observation start time.

\subsection{S6 hardware injections}
LIGO's Science Run 6 (S6) ran from July 2009 to October 2010. During that
observing run,
the LIGO Observatory injected signals for ten mock pulsars
into the data stream by direct excitation of the interferometer mirrors
\citep{2012lscpowerflux}.

We search for each hardware injection
in the raw 64-bit data, from both initial LIGO detectors (Hanford and
Livingston). We then
repeat the search using two different digitisation strategies: a one-bit time
series, representing the most extreme case, and a two-bit time series,
digitised using the optimal thresholds described above.
When digitising, we first apply a bandpass filter that admits only a 2 Hz
frequency band around the injection.
Because some pulsars have significant spin-down, we explicitly include the
injected value of $\dot{f}$ when using \cfs{} to calculate the \fstat{}.
Otherwise, the search
is
performed as described in Section~\ref{ssec:mc}.

The \fstat{} values for the one-, two- and 64-bit searches are presented
for comparison in Table~\ref{tbl:hardwareinj}. The injected pulsars vary widely
in $h_0$, so we do not detect all of them, even in 64-bit data.
For reference, the table also gives the \fstat{} values predicted by the
LALSuite
routine \lalPredictFstat{}, which gives an estimate for $2\mathcal{F}$
based on the observed detector noise and the parameters of the injected signal over the
designated observing time.
The predictions give a sense of whether the signal should be detectable, and
also serve as a check that values of $2\mathcal{F}$ returned by the double-precision search are
sensible.
The results are broadly consistent with Section~\ref{ssec:detrate}, in the sense that
two-bit digitisation reduces $2\mathcal{F}$, for detected pulsars, by
about 10--20 per cent (20--30 per cent in one-bit digitisation),
although pulsar~4 sees a 50 per cent reduction in $2\mathcal{F}$, which occurs
in both digitisation modes tested. In pulsar~3,
when digitising,
the peak \fstat{} appears 80 bins below where it is expected, corresponding to
a reduction in frequency of $5.5\times10^{-4}\,\mathrm{Hz}$ (the peak appears
in the expected bin when processing 64-bit data).

\begin{table}
	\caption
	{One-, two- and 64-bit \fstat{} search results
	for the LIGO Science Run 6 hardware injections,
	using a 20-hr stretch of data starting at GPS time
	$946\,475\,008\,\mathrm{s}$. The
	pulsar indices and injection
	frequencies $f_0$ are those given in \citet{2007abbott}.
	The injections have $\dot{f} \neq 0$, which is treated as a known parameter
	and accounted for directly in $\cfs{}$.
	Column four gives the value of $2\mathcal{F}$ predicted by the LALSuite code
	\lalPredictFstat.
	The detection threshold is
	$2\mathcal{F}_c = 3.9 \times 10^1$
	for false alarm probability $\alpha = 0.01$ over the $1\,\mathrm{Hz}$
	search band. For the injections that are
	detected, marked by *, we report the reduction in $2\mathcal{F}$ when digitising the time series to two-bits using
	optimal thresholds (columns six and seven), and to one-bit
	(columns eight and nine). In pulsar~3, marked by~$\dagger$, the peak in $2\mathcal{F}$
	we report for digitised data is
	shifted by $5.5\times10^{-4}\,\mathrm{Hz}$ from the frequency bin expected
	to contain the signal.}
	\label{tbl:hardwareinj}

		\hspace*{-1.5cm}
		\begin{tabular}{lllllllll} \hline\hline
			Index & $f_0$ (Hz) & $h_0$                & Predicted $2\mathcal{F}$   & $2\mathcal{F}$ (64-bit) & $2\mathcal{F}$ (two-bit) & $2\mathcal{F}_{\mathrm{1\,bit}}/2\mathcal{F}_{\mathrm{64\,bit}}$ & $2\mathcal{F}$ (one-bit) & $2\mathcal{F}_{\mathrm{1\,bit}}/2\mathcal{F}_{\mathrm{64\,bit}}$ \\ \hline
			0 & 265.5          & $2.47\times10^{-25}$ & $4.7$   & $8.5$  & $4.6$ & --      & $13$    & --     \\ 
			1 & 849.1          & $1.06\times10^{-24}$ & $4.8$   & $16$   & $10$  & --      & $0.18$ & --     \\ 
			2* & 575.2          & $4.02\times10^{-24}$ & $99$    & $80$   & $68$  & $0.85$  & $61$    & $0.76$ \\ 
			3* & 108.9          & $1.63\times10^{-23}$ & $430$   & $304$  & $261\dagger$ & $0.86$  & $231\dagger$    & $0.76$ \\ 
			4* & 1430           & $4.56\times10^{-23}$ & $560$   & $539$  & $287$ & $0.53$  & $250$    & $0.46$ \\ 
			5 & 52.8           & $4.85\times10^{-24}$ & $4.9$   & $4.1$  & $5.7$ & --      & $2.7$    & --     \\ 
			6 & 148.7          & $6.92\times10^{-25}$ & $4.9$   & $1.1$  & $2.3$ & --      & $7.5$    & --     \\ 
			7 & 1221           & $2.20\times10^{-24}$ & $9.3$   & $0.48$ & $5.0$ & --      & $4.5$ & --     \\ 
			8* & 194.3          & $1.59\times10^{-23}$ & $440$   & $499$  & $424$ & $0.85$  & $346$    & $0.70$ \\ 
			9 & 763.8          & $8.13\times10^{-25}$ & $4.7$   & $8.6$  & $6.3$ & --      & $8.4$    & --     \\ 
			\hline
		\end{tabular}
\end{table}

\subsection{Software injections}
Ten hardware injections are too few from which to infer reliably a detection
threshold corrected for non-Gaussian noise. To estimate the sensitivity loss
when digitising
actual, non-Gaussian data,
we take a
segment of detector output from S6 and inject synthetic
signals into it.
The synthetic signals are generated in the frequency domain by \mfd{}, which
can use the S6 detector SFTs in lieu of Gaussian noise.
We digitise the time series by
computing the inverse Fourier transform, digitising it, then
computing the
forward Fourier transform.
The SFTs cover the 1 Hz
search band,
so the above procedure
effectively bandpass filters the data.
When digitising in two bits, we use the optimal thresholds.
The search procedure and detection criterion are otherwise the same as described in
Section~\ref{ssec:mc}.
We then
compare the signal strengths needed for detection in one-, two- and 64-bit
data. For the one-bit case, we test digitising in both the time series and the SFTs.

In order to perform a large number of Monte-Carlo--style trials, we exploit the
natural variation in the data in three ways:
varying
right ascension $\alpha$, varying $f_0$, and using
different stretches of S6 data.
We compare 64-bit against digitised data by
taking a signal at the threshold of detectability, when in 64-bit data, and
measuring
how much stronger that signal needs to be, in order to be detectable in
digitised data.
We quantify this using
$B = h_{0,\mathrm{min,digitised}}/h_{0,\mathrm{min,64\,bit}}$,
where $h_{0,\mathrm{min}}$ is the lowest-strength signal that can be detected
(i.e. the lowest $h_0$
with $2\mathcal{F} > 2\mathcal{F}_c$).

\begin{figure}
	\includegraphics[width=0.99\columnwidth]{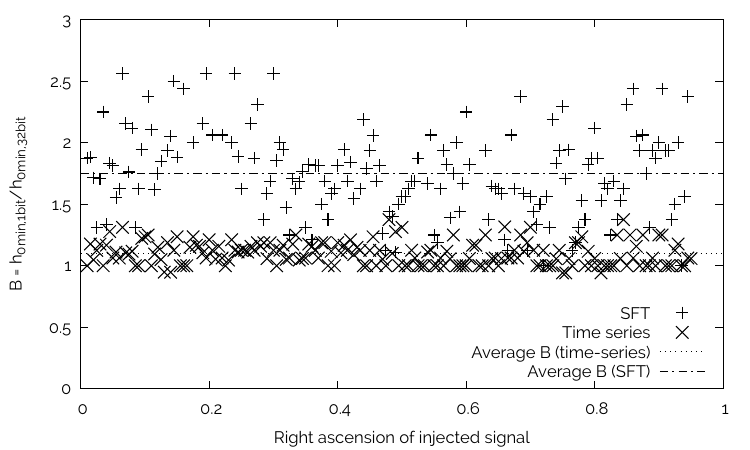}
	\caption
	{
Sensitivity loss factor $B = h_{0,\mathrm{min,1\,bit}}/h_{0,\mathrm{min,64\,bit}}$
for one-bit digitisation
of both the time series and SFT, showing $189$ realisations with different
	$\alpha$ (horizontal axis).
	The crosses show $B$ for one-bit digitised time series, and 
	the pluses show $B$ for $1$-bit digitised SFTs. The average $B$ is shown by
	the dotted (one-bit time series digitisation) and dot-dashed (one-bit SFT
	digitisation) lines, which are the numbers reported in
	Table~\ref{tbl:realdata}.
		}
		\label{fig:skyposb}
\end{figure}

When varying $\alpha$, we use a single
10-hr stretch of data (from both detectors) and try 189 choices of right
ascension $\alpha$ uniformly separated by
$0.005\,\mathrm{rad}$ between $\alpha = 0.01\,\mathrm{rad}$ and $\alpha =
0.95\,\mathrm{rad}$. Other parameters are
given in Table~\ref{tbl:simparams}.
The value of $B$ for each trial is shown in
Figure~\ref{fig:skyposb}, and the average $B$ for those trials is given in
Table~\ref{tbl:realdata}.
In this case,
one-bit time-series digitisation is actually an improvement over the results in
Section~\ref{ssec:detrate}: with real data, we find $B \approx 1.1$
(compared to $B \approx 1.3$ in our trials with Gaussian noise).
In contrast, the results for SFT digitisation are worse, returning
$B = 1.95$ (compared to $B= 1.29$ in trials with Gaussian noise). The two-bit
case improves over the one-bit (time series) case by around eight per cent.

\begin{table}
	\caption
	{Mean of sensitivity loss factor
$B = h_{0,\mathrm{min,digitised}}/h_{0,\mathrm{min,64\,bit}}$,
measuring how much stronger a signal must
	be to be detected in one-bit and two-bit digitised data versus 64-bit digitised data, for
	a signal injected into
	realistic background noise from LIGO Science Run 6.
	Two-bit digitisation is performed using the optimal thresholds.
	We vary sky position, source frequency, and observation starting
	time to produce multiple realisations.
}
	\label{tbl:realdata}
	\begin{center}
		\begin{tabular}{lllll} \hline\hline
			Test & Realisations & $B$ (2-bit time series) & $B$ (1-bit time series) & $B$ (1-bit SFT) \\ \hline
			Sky positions & 189 & 1.10                    & 1.19 & 1.95 \\
			Frequencies & 200   & 1.28                    & 1.31 & 2.61 \\
			Start times & 50    & 1.26                    & 1.32 & 1.99 \\
			\hline
		\end{tabular}
	\end{center}
\end{table}

When varying $f_0$, we use a single 10-hr stretch of data and
inject signals at $0.2\ \mathrm{kHz} \leq f_0 \leq 1\ \mathrm{kHz}$.
The injection frequency $f_0$ is selected by drawing randomly from a uniform
distribution over that range, and other parameters are given in
Table~\ref{tbl:simparams}.
The
loss in sensitivity for the one-bit digitised time series is close to what we
observe in Section~\ref{ssec:detrate}: for realistic data, we find $B \approx
1.2$,
compared to $B \approx 1.3$ for trials on Gaussian noise.
Digitising the time series to two-bits produces only a slight improvement, of
around two per cent.
The SFT digitised
data performs better in this trial than when varying $\alpha$, with
$B \approx 1.4$, although it is still worse than the Gaussian
noise trials (which show $B \approx 1.3$).
We
do not observe a frequency dependence in $B$.

When varying the start time,
we identify 50 non-overlapping ten-hour segments of S6 data.
While not every such segment is complete,
all selected segments have a duty cycle
over 89 per cent, and the \fstat{} can be calculated even in the presence of data
gaps.
We inject a signal, whose parameters are given in Table~\ref{tbl:simparams},
into each segment and search for it in digitised and undigitised data.
The results are similar to those for Gaussian noise for both one- and two-bit
digitisation of the time series.
The
performance of SFTs is worse, with $B \approx 1.99$; that
is, a signal needs to be roughly twice as strong to be detectable.

Overall, the results in Table~\ref{tbl:realdata} are promising: they show that
digitising the time series to one bit yields $1.19 \lesssim B \lesssim 1.32$
(two bits $1.10 \lesssim B \lesssim 1.26$)
for a range of different trials with realistic noise. Digitising the SFTs
yields $1.4 \lesssim B \lesssim 2.5$.
The results suggest that digitisation of the time series works roughly as well
on real data as on
synthetic data,
even without
post-processing the
data to make it
more amenable to digitisation (other than bandpass filtering to the search
band) through procedures such as whitening \citep{2001cuoco}; these procedures
may be necessary when digitising the SFTs.
This is consistent with the result of the analytic calculation in the Appendix,
namely that the autocorrelation function of a one-bit digitised signal is
independent of the power-law exponent for a fat-tailed noise PDF and depends
weakly on the parameters of a thin-tailed noise PDF, e.g, modelled as an
admixture of two Gaussians.

\section{Conclusion}
This paper shows that, when digitising 64-bit interferometer strain data down to
one or two bits, \fstat{}--based gravitational wave searches remain
viable, with a modest sensitivity penalty.
This translates to a factor of 64 or 32
saving in memory, for a sensitivity penalty of around 25 per cent (one-bit
digitisation) or 6.4 per cent (two-bit digitisation).

In any application of digitisation, careful thought should be put into designing the data
pipeline so that digitisation is performed at the most effective place. For
systems that are storage-limited, this might be how the data are stored before
processing. For systems that are bandwidth limited, data might be digitised
before transfer. For systems that are memory-limited, digitisation may be
performed as the data are streamed from storage into memory.
We emphasise that continuous wave searches are not memory-limited normally at
the time of writing, so the case for sacrificing sensitivity to save memory is
marginal, even though the sensitivity loss is modest. However, it is hard to
predict the future of gravitational wave astronomy reliably at such an early
juncture. Future advances in search algorithms, data collection practices, data
volumes, and interferometer designs may create a need for memory savings, just
as occurred historically with the move to baseband recording in pulsar radio
astronomy, e.g. \citet{2000stairs}. If a need for
memory savings does emerge in the future, the results in this paper provide
some preliminary comfort, that digitisation can help, and that the impact on
sensitivity is tolerable.

Places where few-bit digitisation is worth exploring include
I/O-con\-strai\-ned
environments, for example cloud-based searches on commodity commercial
providers (such as Amazon Web Services), or massively distributed
volunteer computing projects such as
Einstein@Home \citep{2010scienceeinsteinathome,2017abbottEAHlowfreq}
through the Berkeley Open Infrastructure for Network Computing (BOINC) project.
Digitisation may also be helpful in novel computing environments that might be
developed in the future, where computing power is plentiful but memory or
transfer is constrained.
A prototypical
example of this evolution is graphics processing units: while they now have adequate memory
bandwidth for the searches described in this paper, early GPUs had
directional, high-latency transfer busses and a low ratio of I/O to compute
resources.
Digitisation to four- or eight-bits also results in negligible sensitivity
loss while delivering memory savings and should be
considered.

While no changes are necessary to the algorithms in
\cfs{} for \fstat{}--based searches,
a new storage format for time series
or SFT files
needs to be developed. This could be integrated into the LIGO Algorithms
Library, or could be a stand-alone wrapper, which converts from the compact
digitised form to the standard format in RAM. Continuous wave search
algorithms that use reduced precision modular arithmetic on a complex
lattice have also been developed
previously \citep{dergachevPC}.


\appendix

\section{One-bit digitisation of non-Gaussian noise}

In this appendix, we investigate theoretically the effect of digitisation on non-Gaussian
noise, by considering a simplified
analogy of the detection process
following
the approach in \citet{1966vanvleck}.
Specifically, we analyse the temporal autocorrelation function of a noisy data
stream as a loose proxy for a matched filter such as the \fstat{}, noting that
the matched filter itself is too challenging to study directly in the analytic
fashion introduced by \citet{1966vanvleck}. We emphasise that the results in
the appendix are illustrative and pedagogical; they should not be applied to
interpret quantitatively the empirical results in Sections~\ref{sec:tests}
and~\ref{sec:nongauss}.
For simplicity, we consider one-bit digitisation implemented according
to Equation~(\ref{eqn:one-bit-sgn}).
For definiteness, we
consider a noise process, whose PDF
is either fat-tailed (power-law) or thin-tailed (sum of two Gaussians).
The analysis can be generalised
directly to multiple bits and PDF tails with other functional forms with
minimal
alterations.

Consider a noisy time series $x(t)$. Define the random variables $X= x(t)$ and
$Y = x(t+\tau)$. The correlation function $r(\tau) = \langle
x(t)x(t+\tau)\rangle$ is given by the covariance of $X$ and $Y$, namely
\begin{align}
	r(\tau)&= \langle XY \rangle
	\label{eqn:a1} \\
		   &= \int_{-\infty}^{\infty} \dee{X} \int_{-\infty}^{\infty} \dee{Y} XY p(X,
	Y),
\end{align}
where $p(X, Y)$ is the joint PDF, i.e., $p(X,Y)\dee{X}\dee{Y}$ equals the
probability of measuring $X \leq x(t) \leq X + \dee{X}$ \textit{and} $Y
\leq x(t+\tau) \leq Y + \dee{Y}$.

Now suppose that the noisy time series is digitised to one bit to give the time
series $\tilde{x}(t) = f[x(t)]$, with
\begin{equation}
	f(X) = \begin{cases}
		1\phantom{-} \qquad \mathrm{for\ } X \geq 0 \\
		-1 \qquad \mathrm{for\ } X < 0.
	\end{cases}
	\label{eqn:a3}
\end{equation}
By the argument leading to equation~(\ref{eqn:a1}), the covariance of the
one-bit digitised signal is given by
\begin{equation}
R(\tau) = \int_{-\infty}^{\infty} \dee{X} \int_{-\infty}^{\infty} \dee{Y}
	f(X) f(Y) p(X, Y).
	\label{eqn:a4}
\end{equation}

We now ask: what is $R(\tau)$ in terms of $r(\tau)$? Specifically, what is
$R(\tau)/r(\tau)$ in the low--signal-to-noise regime $r(\tau) \ll 1$ applicable
to
gravitational wave searches?
Strictly speaking, the relevant figure of merit for gravitational wave searches
is actually the ratio of the
	specific detection statistic (e.g, the \fstat{} in this paper)
for one- and 64-bit digitisation, but the
	latter calculation lies beyond the scope of this paper, as noted above.
	Instead, we use the
correlation function ratio $R(\tau)/r(\tau)$ as a loose analogy for the \fstat{}
ratio. There is no guarantee that the ratio $R(\tau)/r(\tau)$ behaves the same way as
the \fstat{} ratio.
The goal is merely to check for broad, qualitative consistency with the
empirical findings in
Sections~\ref{sec:tests} and~\ref{sec:nongauss},
e.g.\ that the sensitivity loss
does not depend on search parameters in either Gaussian or non-Gaussian
noise, within an analytic framework.

Following \citet{1966vanvleck}, we separate the integral~(\ref{eqn:a4})
with~(\ref{eqn:a3}) into four quadrants to obtain:
\begin{align}
	R(\tau) = &\phantom{-}\int_0^\infty \dee{X} \int_0^\infty \dee{Y} p(X,Y)
	- \int_0^{\infty} \dee{X} \int_{-\infty}^0 \dee{Y} p(X, Y) \notag \\
	&- \int_{-\infty}^{0} \dee{X} \int_0^\infty \dee{Y} p(X, Y) 
	+ \int_{-\infty}^0 \dee{X} \int_{-\infty}^0 \dee{Y} p(X, Y) 
	\label{eqn:a5}
	\\
	=& -1 + 4\int_0^\infty \dee{X} \int_0^\infty \dee{Y} p(X, Y),
	\label{eqn:a6}
\end{align}
where~(\ref{eqn:a6}) follows from~(\ref{eqn:a5}) by using the normalisation condition
$1 = \int_{-\infty}^\infty \dee{X} \int_{-\infty}^\infty \dee{Y} p(X, Y)$.
\citet{1966vanvleck} evaluated~(\ref{eqn:a6}) for the two-dimensional Gaussian
PDF
\begin{equation}
	p(X, Y) = (2\pi)^{-1} |\mathrm{det} \matr{\Sigma}|^{-1/2}
	\mathrm{exp}(-\matr{x}\matr{\Sigma}^{-1}\matr{x}^{\mathrm{T}}/2),
	\label{eqn:app-pdfgauss}
\end{equation}
with
\begin{align}
	\matr{x} &= (\begin{matrix}X & Y \end{matrix}), \\
	\matr{\Sigma} &= \left( \begin{matrix}
\sigma^2 & r \\
	r & \sigma^2
\end{matrix} \right),
\label{eqn:gausspdfdefn}
\end{align}
where $\sigma^2 = \langle X^2 \rangle = \langle Y^2 \rangle$ is the variance,
and $r = \langle X Y \rangle$ is the covariance. The result is
\begin{equation}
	R(\tau) = \frac{2}{\pi} \sin^{-1}\left[\frac{r(\tau)}{\sigma^2}\right].
	\label{eqn:rtaudef}
\end{equation}

In Section~\ref{sec:nongauss} and this appendix, we are interested in how
equation~(\ref{eqn:rtaudef}) generalises for non-Gaussian noise.
Non-Gaussianity takes several forms. One useful limit is a fat-tailed PDF, e.g,
a PDF with a power-law tail, where large noise spikes are more common than a
Gaussian process would predict. This situation occurs whenever the
interferometer is prone to ``glitches'' or experiences elevated levels of
seismic noise \citep{2015aasi}.
Another useful limit is a thin-tailed PDF, where large noise spikes are less
common than a Gaussian process would predict. This can happen when, for
example, the interferometer noise is bistable, switching randomly
between a low and high state, so that the PDF is the sum of a narrow and broad
Gaussian. We consider both limits briefly below.

Consider, for
the sake of definiteness,
a family of fat-tailed PDFs given by the two-dimensional
Student's $t$-distribution
\begin{equation}
	p(X, Y) = (2\pi)^{-1}
	|\det \matr{\Sigma} |^{-1/2} (1 + \nu^{-1} \matr{x} \matr{\Sigma}^{-1}
	\matr{x}^{\mathrm{T}})^{-(2+\nu)/2}
	\label{eqn:studentpdf}
\end{equation}
with
\begin{equation}
	\matr{\Sigma} = \frac{\nu - 2}{\nu} \left( \begin{matrix}
			\sigma^2 & r \\
			r & \sigma^2
	\end{matrix} \right)
	\label{eqn:a12}
\end{equation}
instead of (\ref{eqn:gausspdfdefn}).
Again the covariance satisfies $\langle XY \rangle = r$ and the variance
satisfies $\langle X^2 \rangle = \langle Y^2 \rangle = \sigma^2$. The
covariance is meaningful only for $\nu > 2$. Note that we do \textit{not} claim
that LIGO detector noise obeys the PDF (\ref{eqn:studentpdf}), although it
is a fair approximation at certain epochs, when the interferometer
``glitches''.
Rather, equation~(\ref{eqn:studentpdf}) offers a convenient way to parameterise
fat-tailed, non-Gaussian noise.

Upon substituting~(\ref{eqn:studentpdf}) and~(\ref{eqn:a12})
into~(\ref{eqn:a6}), we obtain the
result
\begin{equation}
	R(\tau) = \frac{2}{\pi}\sin^{-1}\left[\frac{r(\tau)}{\sigma^2}\right].
	\label{eqn:rtaunongauss}
\end{equation}
Remarkably, equation~(\ref{eqn:rtaunongauss}) is identical to
equation~(\ref{eqn:rtaudef}) for all values of $\nu$, i.e., for all values of
the exponent of the power-law tail. The reduction in the covariance is the same
for Gaussian and non-Gaussian noise modelled by the
PDFs~(\ref{eqn:app-pdfgauss}) and~(\ref{eqn:studentpdf}).
A qualitatively similar result can be proved for any PDF which is a function of
$\matr{x}\matr{\Sigma}^{-1}\matr{x}^\mathrm{T}$ by rotating coordinates
in~(\ref{eqn:a6}) to the principal axes of the ellipse
$\matr{x}\matr{\Sigma}^{-1}\matr{x}^{\mathrm{T}} = \mathrm{constant}$.

Now consider a family of thin-tailed PDFs of the form
\begin{equation}
	p(X, Y) = \zeta p_1(X,Y) + (1-\zeta)p_2(XY)
	\label{eqn:thintailpdf}
\end{equation}
with
\begin{align}
	p_i(X, Y) &= (2\pi)^{-1}|\mathrm{det}\matr{\Sigma}_i|^{-1/2}
	\mathrm{exp}(-\matr{x} \matr{\Sigma}_i^{-1}\matr{x}^{\mathrm{T}}/2), \\
	\matr{x} &= (\begin{matrix}X & Y \end{matrix}), \\
	\matr{\Sigma}_i &= \left( \begin{matrix}
\sigma_i^2 & r_i \\
	r_i & \sigma_i^2
\end{matrix} \right).
\end{align}
It is straight-forward to confirm that equation~(\ref{eqn:thintailpdf}) is
thin-tailed for $\sigma_1 < \sigma_2$.
Upon substituting
equation~(\ref{eqn:thintailpdf}) into equation~(\ref{eqn:a6}), we obtain
\begin{equation}
	R(\tau) = \frac{2}{\pi} \left\{
		\zeta\sin^{-1}\left[ \frac{r_1(\tau)}{\sigma_1^2} \right] +
		(1-\zeta)\sin^{-1}\left[ \frac{r_2(\tau)}{\sigma_2^2} \right]
	\right\}.
\end{equation}
\bibliography{references.bib}
\end{document}